\documentclass[sigconf]{acmart}
\usepackage{natbib}
\usepackage[linesnumbered,ruled,vlined]{algorithm2e}
\usepackage{listings}
\usepackage{hyperref}
\usepackage{bbding}
\usepackage{float}
\usepackage{xcolor}
\AtBeginDocument{%
  }
\copyrightyear{2023} 
\acmYear{2023} 
\setcopyright{acmlicensed}\acmConference[WWW '23 Companion]{Companion Proceedings of the ACM Web Conference 2023}{April 30-May 4, 2023}{Austin, TX, USA}
\acmBooktitle{Companion Proceedings of the ACM Web Conference 2023 (WWW '23 Companion), April 30-May 4, 2023, Austin, TX, USA}
\acmPrice{15.00}
\acmDOI{10.1145/3543873.3587311}
\acmISBN{978-1-4503-9419-2/23/04}
\begin{document}
	\title{Reduce API Debugging Overhead via Knowledge Prepositioning}
\author{Shujun Wang}
\email{wangshujun.wsj@alibaba-inc.com}
\orcid{0000-0001-8161-1695}
\affiliation{%
	\institution{Alibaba Group}
	\city{Beijing}
	\country{China}
	\postcode{100000}
}

\author{Yongqiang Tian}
\email{puling.tyq@taobao.com}
\affiliation{%
	\institution{Alibaba Group}
	\city{HangZhou}
	\country{China}
	\postcode{100000}
}

\author{Dengcheng He}
\email{dengcheng.hedc@alibaba-inc.com}
\affiliation{%
	\institution{Alibaba Group}
	\city{HangZhou}
	\country{China}
	\postcode{100000}
}

\renewcommand{\shortauthors}{Wang et al.}


\begin{abstract}
	OpenAPI indicates a behavior where producers offer Application Programming Interfaces (APIs) to help end-users access their data, resources, and services. Generally, API has many parameters that need to be entered. However, it is challenging for users to understand and document these parameters correctly. This paper develops an API workbench to help users learn and debug APIs. Based on this workbench, much exploratory work has been proposed to reduce the overhead of learning and debugging APIs. We explore the knowledge, such as parameter characteristics (e.g., enumerability) and constraints (e.g., maximum/minimum value), from the massive API call logs to narrow the range of parameter values. Then, we propose a fine-grained approach to enrich the API documentation by extracting dependency knowledge between APIs. Finally, we present a learning-based prediction method to predict API execution results before the API is called, significantly reducing user debugging cycles. The experiments evaluated on the online system show that this work's approach substantially improves the user experience of debugging OpenAPIs.
\end{abstract}


\maketitle

\section{Introduction}
Application Programming Interfaces (APIs) play an essential role in modern software development\cite{API,gapbetuseranddesigner,apiuse1,apiuse12}. With the help of APIs, developers can complete their tasks more efficiently\cite{APIEffecient}. OpenAPI indicates a behavior where producers offer Application Programming Interfaces (APIs) to help end-users access their data, resources, and services. Increasingly, many companies have published an ocean of OpenAPIs on the internet\cite{oceanAPI}. For example, as of October 2022, Alibaba Cloud lists over 15560 OpenAPIs for web services . Moreover, more than 60,000 OpenAPIs will be launched soon.

However, "Learning OpenAPIs" is an empirical challenge for end-users of the knowledge gap between API designers and API users\cite{gapbetuseranddesigner}. To bridge this gap, we developed an OpenAPI developer workbench equipped with multiple auxiliary tools to help users learn and employ the OpenAPIs. For instance,API documenration\cite{doc1,docerr3}, API debugging tool and SDK generation tool\footnote{https://github.com/aliyun/darabonba}. Unfortunately, although we provide detailed OpenAPI usage documentation and tools, the success rate of OpenAPI calls is still unsatisfactory\cite{successrate}.

By analyzing the user behavior sequence of the API workbench, we discovered frequent user behavior (Read Documentation $\rightarrow$ Debug APIs). Inspired by the above, in this paper, we address the problem of improving the success rate of API debugging. We conclude that the success rate of API debugging is an important indicator that intuitively reflects the ease of API learning. Crucially, the contributions are outlined as follows:
\begin{itemize}
	\item We design a workbench (https://next.api.alibabacloud.com) to help users learn and use APIs, and we are willing to share all the data from this workbench. 
	\item We extract parameter constraints and rules from massive API call logs. Based on this, we reduced the complexity of some parameters from  \textit{input} to \textit{selection}, e.g., enumeration.
	\item We propose a fine-grained approach to extract dependencies between APIs to ensure that the API is in an executable state.
	\item We present a classification-based API call result prediction method for saving debugging time.
\end{itemize}

\section{Parameter Constraints}
OpenAPIs may have constraints on parameters, such that not all parameters are always required or always optional. The presence of one parameter could cause another parameter to be required. A clear overview of the constraints helps API consumers integrate without additional support and with fewer integration faults. We detect two levels of parameter constraints in our work:
\begin{enumerate}
	\item \textbf{API Level:}  Parameter combination problem, i.e., for the same OpenAPI, the sequence of input parameters may be different since there are optional parameters.
	\item \textbf{Parameter Level:} Mandatory parameter, Enumeration parameter extraction, and Numeric parameter range.
\end{enumerate}

\subsection{Example Extraction} 

The examples of the parameter are essential. An intuitive way to obtain example values of parameters is to extract them from the API call log directly(i.e., Extract parameter values from a correct API call). However, random extracting examples from massive values is insufficient. We believe that the example values of parameters should be universal. For instance, 90\% parameter values are composed of pure English letters, and the examples should be one of the values of these 90\% rather than others. 

This subsection proposes a MapReduce-based Parameter abstraction approach to extract the common features of an ocean of parameter values. 1

\IncMargin{1em}
\begin{algorithm} \SetKwData{Left}{left}\SetKwData{This}{this}\SetKwData{Up}{up} \SetKwFunction{Union}{Union}
	\SetKwFunction{ExtractCommonSubsequence}{ExtractCommonSubsequence}
	\SetKwFunction{TransformAndCompress}{TransformAndCompress}
	\SetKwFunction{LengthComputing}{LengthComputing}
	\SetKwFunction{MergeSubsequence}{MergeSubsequence}
	\SetKwFunction{MergePatterns}{MergePatterns}
	\SetKwFunction{MergeLengths}{MergeLengths}
	\SetKwInOut{Input}{Input}\SetKwInOut{Output}{Output}
	
	\Input{A set of parameter values $V = v_1 \cup v_2 \cup \ldots \cup v_n$} 
	\Output{A common subsequence $s$ \\ A set of parameter patterns $P = p_1 \cup \ldots \cup p_n  $ \\ A set of parameter lengths $L = l_1 \cup l_2 \cup \ldots \cup l_n$}
	\BlankLine 
	
	\textbf{Mapper}
	
	\qquad $s_i \leftarrow$\ExtractCommonSubsequence{$v_i$}\; 
	\qquad $p_i \leftarrow$\TransformAndCompress{$v_i$}\; 
	\qquad $l_i \leftarrow$\LengthComputing{$v_i$}\; 
	
	\textbf{Reducer}
	
	\qquad $s \leftarrow$\MergeSubsequence{$s_i$}\;
	\qquad $P \leftarrow$\MergePatterns{$p_i$}\;
	\qquad $L \leftarrow$\MergeLengths{$l_i$}\;
	\caption{Parameter Abstraction}
	\label{metaparam} 
	\Return $s$, $P$, $L$
\end{algorithm}
\BlankLine
\DecMargin{1em} 

As shown in Algorithm~\ref{metaparam}, we take a two-stage approach: Mapper and Reducer, to extract common parameter features from an ocean of examples. We emphasize three types of parameter features: Component Element Analysis, Element Arrangement Analysis, and Length Analysis.

There are two main stages of parameter abstraction:

\textbf{Mapper:} In the mapper stage, we mainly focus on extracting local features of parameter values. Specifically,
\begin{itemize}
	\item \textbf{Extracting Common Subsequence} will extract the longest common subsequences in the parameter set $v_i$.
	\item \textbf{Transformer And Compress} will translate a specific parameter value into an abstract value. The translation rules are as follows:
	\begin{table}
		\caption{Transform Rules\label{tab:transform}}
		\begin{tabular}{l|l} 
			\hline
			Representation & Description \\
			\hline  
			\textit{z} & The Chinese character \\
			\textit{x} & English lowercase characters  \\
			\textit{X} & English uppercase characters \\
			\textit{d} & Number  \\
			& Other characters reserved\\
			\hline
		\end{tabular}
	\end{table}
	The transformer stage is responsible for abstract parameters. For example, $123$ will be abstracted as $ddd$, and the compressing stage will further translate it into a character $d$.
	\item \textbf{LengthComputing} mainly computes the length information of parameters and their frequency.
\end{itemize}

\textbf{Reducer:} In this stage, the local parameter features are combined to extract the global characteristics of the parameters.

\subsection{Parameter Combination} \label{sec:paramCom}
In order to successfully employ an OpenAPI, there may be different parameter sequences. Take Alibaba Cloud's OpenAPI \textbf{SendSms} (Send Message) as an example, which has two parameter sequences (See Table~\ref{tab:paramSeq}). Note that \textit{Outid} is an optional parameter. 
\begin{table}
	\caption{Parameter Sequences\label{tab:paramSeq}}
	\begin{tabular}{c|c} 
		\hline
		\textbf{	OpenAPI} & \textbf{Parameter Sequences} \\
		\hline 
		SendSms & \scriptsize PhoneNumbers,  SignName, TemplateCode, TemplateParam\\
		SendSms & \scriptsize PhoneNumbers,  SignName, TemplateCode, TemplateParam, OutId  \\
		\hline
	\end{tabular}
\end{table}
\begin{table}[H]
	\caption{Parameter Sequence Rate\label{tab:paramSeqRate}}
	\begin{tabular}{c|c|c} 
		\hline
		\textbf{	OpenAPI} & \textbf{Parameter Sequences} & \textbf{Rate}\\
		\hline 
		SendSms & \scriptsize PhoneNumbers,  SignName, TemplateCode, TemplateParam & 0.7\\
		SendSms & \scriptsize PhoneNumbers,  SignName, TemplateCode, TemplateParam, OutId & 0.3  \\
		\hline
	\end{tabular}
\end{table}

Algorithm~\ref{seqExtaction} outlines the parameter sequence extraction process. We present a MapReduce approach to gather all parameter sequences and compute the frequency. We first extract all parameters from an API request (Line 2). Then, we filter the useless parameters, for example, private or system parameters (Line 3). After this, we can obtain all valuable parameters. Further, we sort these parameters to merge them as keys (Line 4). Finally, we collect all keys (sorted parameter sequence) and compute their frequency (Line 5).
In the reduce stage, we combined the intermediate results (Line 7).
\IncMargin{1em}
\begin{algorithm} \SetKwData{Left}{left}\SetKwData{This}{this}\SetKwData{Up}{up} \SetKwFunction{Union}{Union}
	\SetKwFunction{ExtractParameters}{ExtractParameters}
	\SetKwFunction{FilterParameters}{FilterParameters}
	\SetKwFunction{SortedParametersAsKey}{SortedParametersAsKey}
	\SetKwFunction{MergeKeyCount}{MergeKeyCount}
	\SetKwInOut{Input}{Input}\SetKwInOut{Output}{Output}
	
	\Input{API Requests $R = r_1 \cup r_2 \cup \ldots \cup r_n$} 
	\Output{An Object $O$, the key is the parameter sequence and value is the frequency}
	\BlankLine 
	
	\textbf{Mapper}
	
	\qquad $t_i \leftarrow$\ExtractParameters{$r_i$}\; 
	\qquad $p_i \leftarrow$\FilterParameters{$t_i$}\; 
	\qquad $k_i \leftarrow$\SortedParametersAsKey{$p_i$}\; 
	\qquad $m_i \leftarrow$\MergeKeyCount{$k_i$}\; 
	\textbf{Reducer}
	
	\qquad $O \leftarrow$\MergeKeyCount{$m_i$}\;
	\caption{Parameter Sequences Extraction}
	\label{seqExtaction} 
	\Return $O$
\end{algorithm}
\subsection{Enumeration Parameter}  Enumeration value is an essential parameter type, which reduces the complexity from the input level to the select level. In the case of sufficient computing resources, a simple SQL can complete the enumeration value extraction.

\begin{lstlisting}[language = C++, numbers=left, 
	numberstyle=\tiny,keywordstyle=\color{blue!70},
	commentstyle=\color{red!50!green!50!blue!50},frame=shadowbox,
	rulesepcolor=\color{red!20!green!20!blue!20},basicstyle=\ttfamily]
SELECT COUNT(DISTINCT value) as number,
DISTINCT value as values
FROM An ocean of parameter values 
WHERE number < 20
\end{lstlisting}

Where 20 is a hyperparameter.
In the case of insufficient computing resources, suppose we have obtained a parameter $p_i$ of OpenAPI $a_i$ is enumerable, and the enumeration value set is $v_ i$. After that, we process a small part of the data every day and get an enumeration value set of $v_ j$, and we only need to guarantee | $v_i \cup v_j$|< 20.
\section{API Dependencies Computing}
We propose a fine-grained approach to compute the parameter-level OpenAPI relevance in this subsection. We have noticed that the input parameter of one OpenAPI is often the output parameter of some other OpenAPIs.
\begin{table}[htbp]
	\begin{minipage}[t]{0.45\linewidth}
		\caption{SendSms\label{tab:sendsms}}
		\centering
		\begin{tabular}{c|c} 
			\hline
			\small	Input  & \small Output \\
			\hline 
			\scriptsize PhoneNumbers & \scriptsize Code \\
			\scriptsize \color{red}{SignName} & \scriptsize Message  \\
			\scriptsize TemplateCode &\scriptsize BizId \\
			\scriptsize TemplateParam &\scriptsize  RequestId \\
			\scriptsize OutId &\\
			\hline
		\end{tabular}
	\end{minipage}
	\begin{minipage}[t]{0.45\linewidth}
		\caption{ AddSmsSign\label{addsmssign}}
		\centering
		\begin{tabular}{c|c} 
			\hline
			\small Input  & \small Output \\
			\hline 
			\scriptsize SignName & \scriptsize Code \\
			\scriptsize Remark & \scriptsize Message  \\
			\scriptsize SignFileList & \scriptsize \color{red}{SignName} \\
			\scriptsize TemplateParam & \scriptsize RequestId \\
			\hline
		\end{tabular}
	\end{minipage}
\end{table}

Table~\ref{tab:sendsms}, \ref{addsmssign} exhibits the input parameters and output parameters of two OpenAPIs (i.e., SendSms, AddSmsSign). The output parameter \textit{SignName} of AddSmsSign is also the input parameter of SendSms. Regarding query semantics, if the user wants to use a \textit{SignName}, it should be created first.

We design a generation and ranking approach to compute this co-occurrence.
\begin{itemize}
	\item \textbf{Candidate Generation:}  For any OpenAPI input parameter $p_i$, we output a set of OpenAPIs $C$, whose output parameter contains $p_i$.
	\item \textbf{Ranking:} Given an OpenAPI $a_i$ and an input parameter $p_i$ of $a_i$, the task of this stage is to rank the OpenAPI in $C$, and then recommend a subset of $C$ to end-users. Given an OpenAPI $c$ of $C$, the score of $s(a_i, p_i, c_i)$ as follows:
	\begin{equation}
		\label{eq:rankscore}
		s(a_i, p_i, c) =\frac{  \alpha sim(a_i, c) \cdot  \beta sim(p_i, c) \cdot  \sigma s(a_i, c )}{c_i + c_o }
	\end{equation}
	Where $\alpha, \beta, \sigma$ is the hyperparameter, $sim(x, y)$ means the string similarity between $x$ and $y$, and $s(x, y)$ is the OpenAPI relevance between $x$ and $y$.
\end{itemize}

\section{Char-Level Error  Prediction}
In this subsection, we mainly discuss how to predict the result of an API request before the request is executed. We reduce the result prediction task to a multi-classification job to solve this problem. The categories are various Error Codes and the $\langle$right$\rangle$ label, representing the correct execution.

An example of \textit{SendMessage} as follows:
\\
\begin{lstlisting}[language = C++, numbers=left, 
	numberstyle=\tiny,keywordstyle=\color{blue!70},
	commentstyle=\color{red!50!green!50!blue!50},frame=shadowbox,
	rulesepcolor=\color{red!20!green!20!blue!20},basicstyle=\ttfamily]
{
 "PhoneNumbers":"186xxx9602",
 "SignName":"hanxing",
 "TemplateCode":"SMS_209470795"
}
\end{lstlisting}

The parameter value does not have actual meaning, such as the value of TemplateCode, i.e., SMS\_209470795. 
This paper explores treating text as a raw signal at the character level and applying temporal (one-dimensional) ConvNets to solve it.

\subsection{Key Modules}
The main component is the temporal convolutional module, which computes a 1-D convolution. Suppose we have a discrete input function $g(x) \in[1, l] \rightarrow \mathbb{R}$ and a discrete kernel function $f(x) \in[1, k] \rightarrow \mathbb{R}$ The convolution $h(y) \in[1,\lfloor(l-\dot{k}) / d\rfloor+1] \rightarrow \mathbb{R}$ between $f(x)$ and $g(x)$ with stride $d$ is defined as 
\begin{equation}
	h(y)=\sum_{x=1}^{k} f(x) \cdot g(y \cdot d-x+c)
\end{equation}

Where $c=k-d+1$ is an offset constant, just as in traditional convolutional networks in vision, the module is parameterized by a set of such kernel functions $f_{i j}(x)(i=1,2, \ldots, m$ and $j=$ $1,2, \ldots, n)$ which we call weights, on a set of inputs $g_{i}(x)$ and outputs $h_{j}(y) .$ We call each $g_{i}$ (or $h_{j}$ ) input (or output) features, and $m$ (or $n$ ) input (or output) feature size. The outputs $h_{j}(y)$ is obtained by a sum over $i$ of the convolutions between $g_{i}(x)$ and $f_{i j}(x)$

One key module that helped us to train deeper models is temporal max-pooling. It is the 1 -D version of the max-pooling module used in computer vision [2]. Given a discrete input function $g(x) \in$ $[1, l] \rightarrow \mathbb{R}$, the max-pooling function $h(y) \in[1,\lfloor(l-k) / d\rfloor+1] \rightarrow \mathbb{R}$ of $g(x)$ is defined as
$$
h(y)=\max _{x=1}^{k} g(y \cdot d-x+c)
$$

Where $c=k-d+1$ is an offset constant, this pooling module enabled us to train ConvNets deeper than six layers, where all others fail.
The non-linearity used in our model is the rectifier or thresholding function $h(x)=\max \{0, x\}$, which makes our convolutional layers similar to rectified linear units (ReLUs). The algorithm used is stochastic gradient descent (SGD) with a minibatch of size 128, using momentum $0.9$ and initial step size $0.01$, which is halved every three epochs for ten times. 

\subsection{Character quantization}
Our models accept a sequence of encoded characters as input. The encoding is done by prescribing an alphabet of size $m$ for the input language and then quantizing each character using 1 -of- $m$ encoding (or "one-hot" encoding). Then, the sequence of characters is transformed to a sequence of such $m$ sized vectors with fixed length $l_{0}$. Any character exceeding length $l_{0}$ is ignored, and any characters not in the alphabet, including blank characters, are quantized as all-zero vectors. The character quantization order is backward. The latest reading on characters is always placed near the beginning of the output, making it easy for fully connected layers to associate weights with the latest reading.
The alphabet used in our models consists of 96 characters, including 52 english letters (Case Sensitive), 10 digits, 33 other characters. 
\begin{lstlisting}[language = C++, numbers=left, 
	numberstyle=\tiny,keywordstyle=\color{blue!70},
	commentstyle=\color{red!50!green!50!blue!50},frame=shadowbox,
	rulesepcolor=\color{red!20!green!20!blue!20},basicstyle=\ttfamily]
 abcdefghijklmnopqrstuvwxyz 
 ABCDEFGHIJKLMNOPQRSTUVWXYZ  
 0123456789 
 -,;/!?:/\|_@#$%^&*~`+-=<>(){}
\end{lstlisting}

\section{Evaluation}
In this section, we briefly introduce the metrics used in subsequent experiments. $sr$ represents the success rate of OpenAPI calls.
\begin{equation}
	\label{eq:sr}
	sr = \dfrac{|Call~Success|}{|Call~Number|}
\end{equation}
\subsection{Overall Evaluation}
\begin{table}[htpb]
	\caption{Success Rate Comparision\label{tab:ex:sr}}
	\begin{tabular}{l|ccc|cc}
		\hline Model & Period 1&  Period 2 &   Period 3 &  Period 4  & Period 5 \\
		\hline Baseline & $23.2\%$ & $24.5\%$ &  $25.3\%$ & $\XSolid$ & $\XSolid$ \\
		Base-DE& $\XSolid$ & $\XSolid$ &  $\XSolid$ & $49.7\%$ & $53.2\%$ \\
		\hline
	\end{tabular}
\end{table}
The core goal of this work is to reduce the cost of users learning API and improve the API success rate. 

Table~\ref{tab:ex:sr} shows that the proposed optimization strategies (w.r.t Debugging Experience Optimization and Documentation experience Optimization) achieved better $sr$ scores than pure baseline. The Baseline model refers to the workbench that does not add optimization solutions. Base-DE means that it increased the API debugging strategy. One period means one week. Our experiments are based on real statistics from online platforms. In Period 4, we launched the API parameter constraints and rules optimization strategy, and in Period 5, we launched the prediction optimization strategy.

\subsection{Prediction Evaluation}

We designed 2 ConvNets - one large and one small. They are 9 layers deep with 6 convolutional layers and 3 fully-connected layers. 
The input has several features equal to 96 due to our character quantization method, and the input feature length is 1014 . It seems that 1014 characters could already capture most of the texts of interest. We also insert 2 dropout modules in between the 3 fully-connected layers to regularize, and the dropout probability is $0.5 .$

We initialize the weights using a Gaussian distribution. The mean and standard deviation for initializing the large model is $(0,0.02)$ and the small model $(0,0.05)$.

\begin{table}
	\caption{Fully-connected layers used in our experiments. The problem determines the number of output units for the last layer. For example, for a 10 -class classification problem, it will be $10 .$}
	\begin{tabular}{ccc} 
		\toprule
		Layer & Output Units Large & Output Units Small \\
		\hline 7 & 2048 & 1024 \\
		8 & 2048 & 1024 \\
		9 &\multicolumn{2}{c}{Depends on the problem} \\
		\bottomrule
	\end{tabular}
\end{table}

For different problems, the input lengths may be different (for example, in our case, $l_{0}=1014$ ), and so are the frame lengths. From our model design, it is easy to know that given input length $l_{0}$, the output frame length after the last convolutional layer (but before any of the fully-connected layers) is $l_{6}=\left(l_{0}-96\right) / 27$. This number multiplied with the frame size at layer 6 will give the input dimension the first fully-connected layer accepts.
\begin{table}[H]
	\caption{Precision of all the models\label{tab:precision}}  
	\begin{tabular}{lc} 
		\toprule
		Model & Precision\\
		\hline 
		Lg. Full Conv & $80.61\%$ \\
		Sm. Full Conv & $85.36\%$ \\
		\bottomrule
	\end{tabular}
\end{table}
Table~\ref{tab:precision} shows the reliable results of our char-level models.

\section*{CONCLUSIONS}
APIs have been widely used but are often poorly learned since many parameters need to be entered. A direct proof is that the success rate of API debugging is extremely low (<37\%). This phenomenon causes the loss of a large number of potential users. To reduce the overhead of learning APIs, we developed an API workbench to help users understand and debug APIs; based on this workbench, we collected massive user debug logs and conducted exploratory research. The practice has shown that parameter limitation, rule mining, and parameter range limitation are practical solutions to reduce the cost of learning APIs for users. In addition, helping users know the API's possible execution results in advance is a remarkably effective solution for improving the API debugging experience.
\section*{Acknowledgments}

This work was supported by Alibaba Group through Alibaba Innovative Research Program.

\bibliographystyle{ACM-Reference-Format} \nocite{1}
\bibliography{reference}

\end{document}